\documentclass[preprintnumbers, pra, showpacs, floatfix,twocolumn,
preprintnumbers, letterpaper, superscriptaddress]{revtex4}
\usepackage{amsfonts}
\usepackage{amsmath}
\usepackage{amssymb,epsf}
\usepackage{latexsym}
\usepackage{graphicx,epsfig}
\usepackage{amssymb}
\usepackage{subfigure}
\usepackage[colorlinks=true,citecolor=blue,linkcolor=blue,urlcolor=black]{hyperref}
\usepackage{epstopdf}
\usepackage{color}

\def\be{\begin{equation}}
\def\ee{\end{equation}}
\def\ba{\begin{eqnarray}}
\def\ea{\end{eqnarray}}
\def\la{\langle}
\def\ra{\rangle}

\begin{document}
\title{Phase transition in a noisy Kitaev toric code model}
\author{Mohammad Hossein Zarei}
\email{mzarei92@shirazu.ac.ir}
\affiliation{Physics Department, College of Sciences, Shiraz University, Shiraz 71454, Iran}
\author{Afshin Montakhab}
\email{montakhab@shirazu.ac.ir}
\affiliation{Physics Department, College of Sciences, Shiraz University, Shiraz 71454,
Iran}
\begin{abstract}
It is well-known that the partition function of a classical spin
model can be mapped to a quantum entangled state where some
properties on one side can be used to find new properties on the
other side. However, the consequences of the existence of a
classical (critical) phase transition on the corresponding quantum
state has been mostly ignored. This is particularly interesting
since the classical partition function exhibits non-analytic
behavior at the critical point and such behavior could have
important consequences on the quantum side. In this paper, we
consider this problem for an important example of Kitaev toric
code model which has been shown to correspond to the
two-dimensional (2D) Ising model though a duality transformation.
Through such duality transformation, it is shown that the
temperature on the classical side is mapped to bit-flip noise on
the quantum side. It is then shown that a transition from a
coherent superposition of a given quantum state to a non-coherent
mixture corresponds exactly to paramagnetic-ferromagnetic phase
transition in the Ising model. To identify such a transition
further, we define an order parameter to characterize the
decoherency of such a mixture and show that it behaves similar to
the order parameter (magnetization) of 2D Ising model, a behavior
that is interpreted as a robust coherency in the toric code model.
Furthermore, we consider other properties of the noisy toric code
model exactly at the critical point. We show that there is a
relative stability to noise for the toric code state at the
critical noise which is revealed by a relative reduction in
susceptibility to noise. We close the paper with a discussion on
connection between the robust coherency as well as the critical
stability with topological order of the toric code model.
\end{abstract}
\maketitle
\section{Introduction}
Among the well-known connections from classical statistical
mechanics to quantum information theory
\cite{Geraci2008,Lidar1997,Geraci2010,Somma2007,Verstraete2006,Dennis2002,Katzgraber2009,ent2006,mont2010,eis17,termo},
a fascinating correspondence between partition functions of
classical spin models and quantum entangled states has attracted
much attention \cite{Nest2007,algor,durmar}. In 2007, it was shown
that the partition function of a classical spin model can be
written as an inner product of a product-state and an entangled
state \cite{Nest2007}. Such mappings led to a cross-fertilization
between quantum information theory and statistical mechanics
\cite{gemma,Cuevas2011}. Specifically, it has been shown that
measurement-based quantum computation (MBQC) on quantum entangled
states \cite{mqc, mbqc} is related to computational complexity of
classical spin models \cite{Bravyi2007,Bombin2008}. In this way, a
new concept of the completeness was defined where the partition
function of a classical spin model generates the partition
function of all classical models
\cite{Nest2008,Vahid2012b,Cuevas2009,xu,Vahid2012,yahya}, see also
\cite{science,cub} for recent developments in this direction. Most
such studies were based on a specific mappings between
classical-quantum models.  However, we have recently introduced a
canonical relation as a duality mapping where any given CSS
quantum state can be mapped, via hypergraph representations, to an
arbitrary classical spin model \cite{zare18}.

On the other hand, the problem of phase transition in classical
spin models has attracted much attention in the past and is
therefore a well-studied phenomenon \cite{stan}. Simply, in the high
temperature phase such models exhibit no net magnetization due to
the symmetric behavior of dynamical variables.  Upon decreasing
the temperature, this symmetry is spontaneously broken at a
critical temperature $T_{cr}$, and a non-zero magnetization
appears in the system. The behavior of such systems at the
critical point are characterized by non-analytic properties of the
leading thermodynamics functions such as magnetic susceptibility.
Such non-analytic behavior is characterized by a set of critical
exponents which fully identify the symmetry-breaking property (or
universality class) of the particular phase transition
\cite{pathr}.

Now, since there is a correspondence between such classical spin models
 and entangled quantum states, one would have
to wonder what the consequences of such phase transitions are on
the quantum states.  It is our intention to take a step in this
direction by considering the well-known ferromagnetic phase
transition in a 2D Ising model and its consequences on the Kitaev
toric code (TC) \cite{Kitaev2003}, which we have previously shown
to be related via a duality mapping \cite{zare18}. The TC state is
of particular interest since it has a topological order
\cite{wen,wen2} with a robust nature
\cite{rob1,rob2,zare16,zareiprb} as well as an important
application in quantum error
correction\cite{errork,errorw,errorb}. On the other hand the 2D
Ising model is a well-known model in standard statistical
mechanics which allows an exact solution.  Therefore, one can hope
that exploration of such a mapping between the partition function
of the Ising model and the TC code can open an avenue for many
possible studies related to topological properties of the TC
state.

Subsequently, we consider the TC in the presence of an independent
bit-flip noise where the Pauli operators $X$ are applied to each
qubit with probability $p$. We consider the effect of such a noise
in a coherent superposition of two specific quantum states in the
TC. Then we define an order parameter that can characterize
decoherency of the above quantum state. Interestingly, we show
that such an order parameter is mapped to the magnetization of the
Ising model. Therefore, we conclude that there is a phase
transition from a coherent superposition to a non-coherent mixture
of the above quantum state at a critical probability of $p_{cr}$
corresponding to critical temperature of 2D Ising model. We
interpret such a behavior as a robust coherency in the TC model.
On the other hand, it is well-known that criticality can be
marked by interesting behavior at the transition point. Therefore,
we define a quantity as susceptibility to noise in the noisy TC
model and we show that, at the critical noise of $p_{cr}$, the
susceptibility shows a relative reduction which is indicative of
critical stability which has been pointed out before
\cite{zare18}.

This Article is structured as follows:  In Sec.(\ref{s1}), we give
an introduction to TC model including its ground state and
excitations. In Sec.(\ref{s2}), we review the duality mapping from
the partition function of 2D Ising model to the TC state and
specifically show how such a problem is related to a TC state
under a bit-flip noise. In Sec.(\ref{s3}), we provide our main
result where we introduce a decoherence process for a coherent
superposition of two quantum state in the TC and we find a
singular phase transition to non-coherent phase which is mapped to
the ferromagnetic phase transition in the 2D Ising model. In
Sec.(\ref{s4}), we introduce a susceptibility to noise which
reveals a relative (critical) stability of the toric code state
against bit-flip noise at the transition point.

\section{Review of Kitaev TC model}\label{s1}

Toric code (TC) is the first well-known topological quantum code
which was introduced by Kitaev  in 2003 \cite{Kitaev2003}. Since
we will consider behavior of this model under noise, here we give
a brief review on the TC model which is specifically defined on a
2D square lattice with periodic boundary condition (i.e. on a
torus). To this end, consider a $L\times L$ square lattice where
qubits live on edges of the lattice. Corresponding to each vertex
and face of the lattice, two stabilizer operators are defined in
the following form:
\begin{equation}\label{dd1}
B_f =\prod_{i\in \partial f}X_i ~~,~~ A_v =\prod_{i\in v}Z_i
\end{equation}
where $i\in \partial f$ refers to all qubits living on edges of
the face $f$ and $i \in v$ refers to qubits living on edges
incoming to the vertex $v$, see Fig.(\ref{k1}-a). The above operators are
 in fact generators of the stabilizer group of the TC where each product of
them is also a stabilizer. For example, if we represent each face operator
of $B_f$ as a loop around the boundary of the corresponding face, each
 product of them will also have a loop representation. In this way,  corresponding
to each kind of loop in the lattice, there will be an $X$-type stabilizer.

  On a torus topology, there are two relations between these operators in the
form of $\prod_{f}B_f =I$ and $\prod_{v}A_v =I$ where
$I$ refers to the Identity operator. In this way, the number of
independent stabilizers is equal to $2L-2$. By the fact that  $[A_v
,B_f]=0$, it is simple to show that the following state is an
eigenstate of all face and vertex operators with eigenvalue $+1$:

$$
|K\ra = \frac{1}{\sqrt{2^{(|f|-1)}}}\prod_{f}(I+B_f )|0\rangle ^{\otimes 2L}
$$

\begin{equation}\label{dd2}
=\frac{1}{\sqrt{2^{(|v|-1)}}}\prod_{v}(I+A_v )|+\rangle ^{\otimes 2L}
\end{equation}
where $\prod_f$ refers to product of all independent face
operators and $\prod_v$ refers to product of all independent
vertex operators. $|f|$ and $|v|$ refer to the number of faces and
vertices, respectively. $|0\ra$ and $|+\ra
=\frac{1}{\sqrt{2}}(|0\ra +|1\ra ) $ are positive eigenstates of
Pauli operators $ Z $ and $X $, respectively. The stabilizer space of
the toric code is four-fold degenerate and thus there are three
other stabilizer states which are generated by non-local
operators. In fact, one can consider two non-trivial loops around
the torus in two different directions, see Fig.(\ref{k1}-b). Then
two operators corresponding to non-trivial loops $\gamma$ and $\gamma'$ are
defined in the following form:
\begin{figure}[t]
\centering
\includegraphics[width=8cm,height=10cm,angle=0]{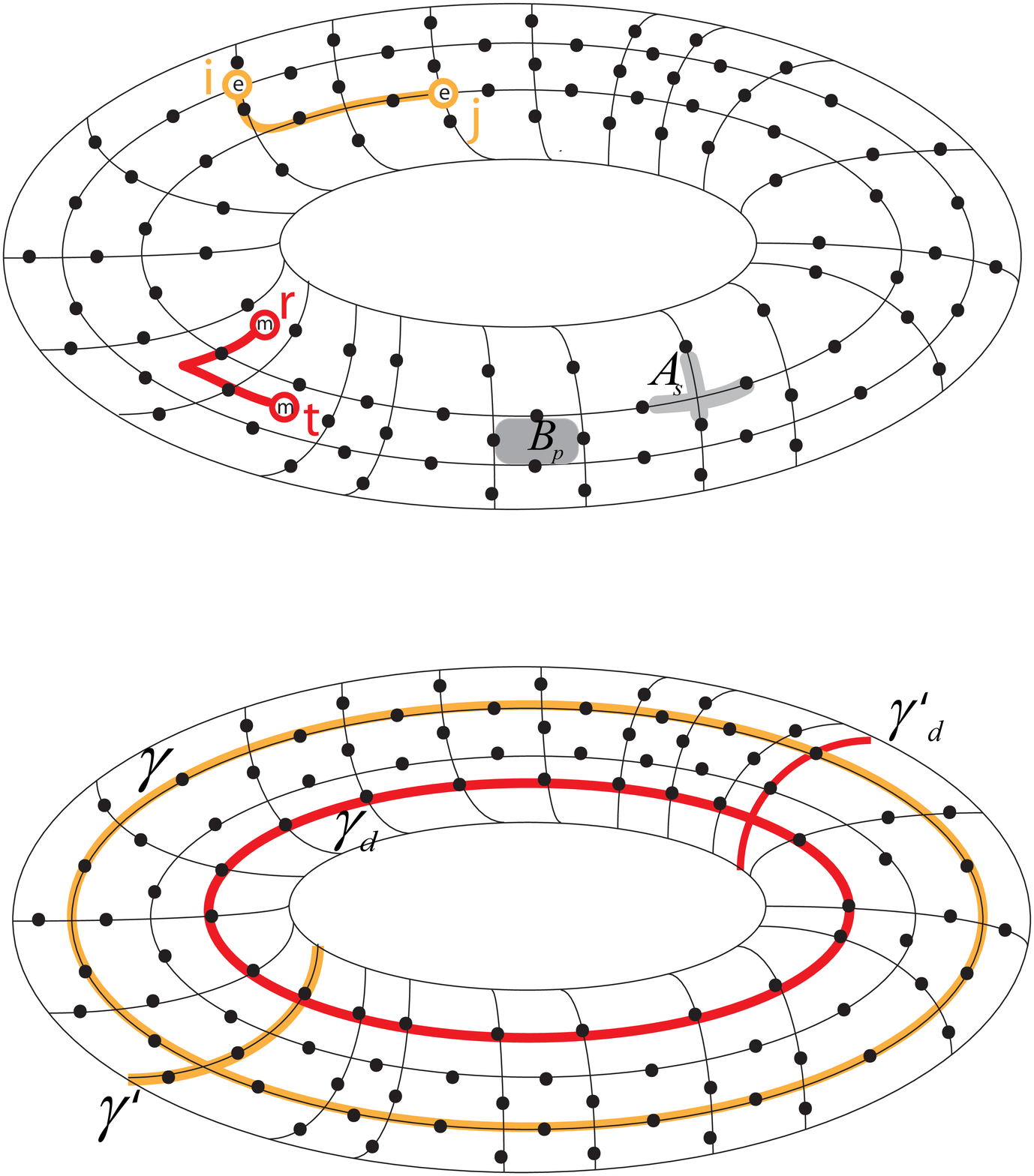}
\caption{(Color online) a) A 2D square lattice on a torus where
qubits live on edges of the lattice. Each vertex (face) operator
$A_v$ ($B_f$) is defined corresponding to each vertex (face) of
the lattice. b) There are two different direction on the lattice
where two non-trivial loops can be defined. The two non-trivial
loops on the edges of the lattice are denoted by $\gamma$ and $\gamma'$
while on the dual lattice they are denoted by $\gamma_d$ and $\gamma'_d$. }
\label{k1}
\end{figure}
\begin{equation}\label{eq1}
\Gamma _x =\prod_{i \in \gamma}X_i ~~~,~~~\Gamma'_{x}=\prod_{i\in \gamma'}X_i
\end{equation}
where $i \in \gamma $ and $i\in \gamma'$ refers to all qubits living in these loops. In this way, the following four quantum states
will be the bases of stabilizer space:
\begin{equation}\label{ewrq}
|\psi_{\mu ,\nu}\ra = (\Gamma _{x})^{\mu} (\Gamma' _{x})^{\nu} |K\ra ~,
\end{equation}
where $\mu, \nu=0, 1$ refer to exponent of non-trivial loop
operators. In addition to the above non-local operators, there are
also two other non-local operators constructed by $Z$ operators.
Such operators correspond to two loops $\gamma_d$ and $\gamma'_d$ around the
torus on dual lattice in the form of
$\Gamma_{z}=\prod_{i\in \gamma_d}Z_i$ and
$\Gamma'_{z}=\prod_{i\in \gamma'_d}Z_i$, see Fig.(\ref{k1}-b).
One can check that these operators can characterize four different
bases of stabilizer space where expectation values of these
operators are different for the bases (\ref{ewrq}).

Another important property of the TC state is related to
excitations of the model. To this end, consider two vertices of
the lattice denoted by $i$, $j$ where a string, denoted by
$S_{ij}$, can connect these two vertices, see Fig.(\ref{k1}-a).
Then we apply the Pauli operators $X$ on all qubits belonging to
the string $S_{ij}$ where we denote the corresponding string
operator by $S_{ij}^{x}$. It is clear that such an operator
commutes with all vertex operators $A_v$ instead of $A_i$ and
$A_j$, which are the two end-points of the string $S_{ij}$. Such
an excited state can also  be interpreted as two charge anyons at
the two end-points of $S_{ij}$. Charge anyons are generated as
pairs and one can move one of them in the lattice by applying a
chain of Pauli operators. Furthermore, we can also define string
operators of the Pauli operators $Z$. To this end, consider two
faces $r$ and $t$ where a string $S_{rt}$ can connect them, see
Fig.(\ref{k1}-a). One can define a string operator $S_{rt}^{z}$
which is a product of $Z$ operators on the $S_{rt}$. Similarly,
such an operator does not commute with two face operators $B_r$
and $B_t$ at the two end-points of the $S_{rt}$, and it is
interpreted as two flux anyons at the end-points of the $S_{rt}$.

\begin{figure}[t]
\centering
\includegraphics[width=5cm,height=4.5cm,angle=0]{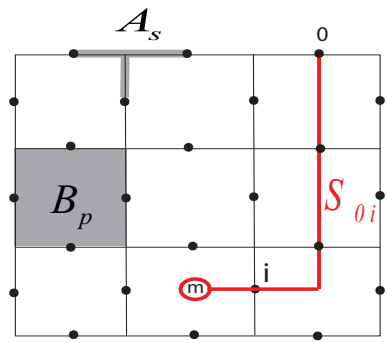}
\caption{(Color online) TC model on a 2D square lattice with open
boundary. The vertex operators corresponding to vertices on the
boundary are three-local. If a string start in the boundary the
corresponding operator generates only one flux anyon at the
end-point of the string in the lattice.} \label{k2}
\end{figure}
We should emphasize that TC model can also be defined on other
lattices with different topologies. The most important difference
between different topologies is related to degeneracy of the
stabilizer space. Specifically in this paper, we consider a two
dimensional square lattice with an open boundary condition, see
Fig.(\ref{k2}). Vertex and face operators are defined similar to
Eq.(\ref{dd1}). However, note that vertex operators corresponding
to vertices of the boundary of the lattice are three-body local.
It is simple to check that unlike the TC on torus, there is only
one constraint on vertex operators, $\prod_{v}A_v =I$, and no
constraint on face operators. In this way, the degeneracy of
stabilizer space will be equal to two. It is also interesting to
consider excitation of this model. Unlike the TC on a torus, here
one can find flux anyons in odd numbers. In fact if we apply a $Z$
operator on a qubit on the boundary of the lattice it will only
generate one flux anyon in the neighboring face.  The other flux
anyon always lives on the boundary of the lattice. In the other
words, the corresponding string operator has two end-points with
one on the boundary and another inside the lattice. We denote such
a string operator by $S_{0i}$ where $0$ refers to a qubit on the
boundary and $i$ refers to a qubit inside the lattice, see
Fig.(\ref{k2}).

\section{Mapping the Ising model to a noisy TC model}\label{s2}
It is well-known that the partition function of a classical spin
model can be mapped to an inner product of a product state and an
entangled state \cite{Nest2007}. We have recently provided
such a mapping using a duality transformation for CSS states which
are mapped to classical spin models \cite{zare18}. In this
section, we review such mapping between the TC state and 2D Ising
model. We also show how a change of variable allows the
temperature in the Ising model to be transformed to bit-flip noise
in the TC state. To this end, we start with the partition function
of a 2D Ising model which is defined on a 2D square lattice with
an open boundary condition where we suppose all spins in the
boundary are fixed to value of $+1$. The partition function will
be in the following form:
\begin{equation}\label{aq1}
\mathcal{Z}=\sum_{\{\sigma_i\}}e^{\beta J \sum_{\la i,j\ra}\sigma_i \sigma_j }
\end{equation}
where $\sigma_i=\{\pm 1\}$ refers to spin variables which live on
vertices of the lattice which we call vertex spins, $J$ refers to
coupling constants and $\beta=\frac{1}{k_B T}$. Now, we define new
spin variables $\xi_l$ which live on edges of the square lattice
which we call edge spins. In Fig.(\ref{k4}), we show these
new spins by green circles. We also define the value of each edge
spin $\xi_l$ in the form of $\xi_l =\sigma_i \sigma_j$ where
$\sigma_i$ and $\sigma_j$ are two vertex spins which live on two
end-points of the edge $l$. In the next step, we re-write the
partition function of Eq.(\ref{aq1}) in terms of the edge spins
$\xi_l$ in the following from:
\begin{figure}[t]
\centering
\includegraphics[width=5cm,height=4.5cm,angle=0]{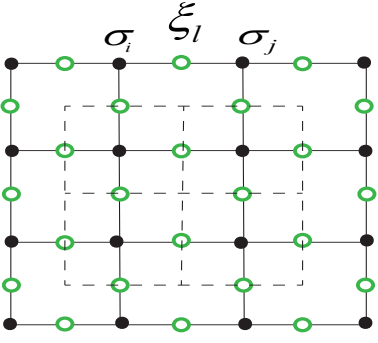}
\caption{(Color online) Spin variables of the 2D Ising model are
denoted by black circles. Corresponding to each edge of the
lattice a new spin variable is defined and is denoted by green
circle. Dashed lattice is the dual of the initial lattice where
faces and vertices of the initial lattice correspond to vertices
and faces of the dual lattice, respectively.} \label{k4}
\end{figure}
\begin{equation}\label{34}
\mathcal{Z}=\sum_{\{\xi_l \}}e^{\beta J \sum_{l}\xi_l } \prod_{f}\delta(\prod_{l \in \partial f}\xi_l)
\end{equation}
where $l \in \partial f$ refers to edges around the face of $f$
and we have added delta functions corresponding to each face of
the lattice in order to satisfy constraints between edge spins. In
other words, since $\xi_l =\sigma_i \sigma_j$, it is clear that
the product of spin variables corresponding to each face of the
lattice will be equal to one. We should emphasize that one can
find another representation for constraints based on dual lattice
which is the same as square lattice. As it is shown in
Fig.(\ref{k4}), each face of the square lattice corresponds to a
vertex of the dual lattice. In this way, constraints in the form
of $\prod_{f}\delta(\prod_{l \in \partial f}\xi_l)$ in
Eq.(\ref{34}) can be replaced by $\prod_{v_d}\delta(\prod_{l \in
v_d}\xi_l)$ where $v_d$ refers to a vertex of the dual lattice.
Here, we use such dual representation for finding quantum
formalism of 2D Ising model.

Now, we re-write each delta function in the form of
$\delta(\prod_{l \in  v_d}\xi_l)=\frac{1+\prod_{l \in
v_d}\xi_l}{2}$. Next, the partition function will be written in a
quantum language in the following form:
\begin{equation}\label{aq2}
\mathcal{Z}=\la \alpha | G\ra
\end{equation}
where $|\alpha \ra =(e^{\beta J} |0\ra +e^{-\beta
J}|1\ra)^{\otimes N_d}$ and $|G\ra =\prod_{v_d}\frac{I+\prod_{l
\in v_d}Z_l}{2}(|0\ra +|1\ra)^{\otimes N_d}$ where $N_d$ is the
number of edges of the dual lattice. By comparison with
Eq.(\ref{dd2}) and by the fact that $\prod_{l \in  v_d}Z_l$ is in
fact a vertex operator $A_v$ on the dual lattice, it is clear that
$|G\ra$ is the same as the toric code state $|K\ra$ on the dual
lattice up to a correction in normalization factor. Finally, the
partition function will be in the form of:
\begin{equation}\label{aq3}
\mathcal{Z}=\sqrt{2^{|f_d|}} \la \alpha | K\ra,
\end{equation}
where $|f_d|$ is the number of faces of the dual lattice. In this
way, the partition function of the 2D Ising model on a square
lattice is related to a TC state on the dual lattice with qubits
which live on the edges.

Now, let us define a new quantity $p$ which is related to
Boltzmann weight in the form of $p=\frac{e^{-2\beta J}}{1+e^{-2\beta
J}}$. Since $\beta J$ is a quantity between zero to infinity, it
is concluded that $0<p<\frac{1}{2}$. In terms of this new
quantity, the partition function can be re-written in the
following form:
\begin{equation}\label{ds}
\mathcal{Z}=\frac{1}{[p(1-p)]^{\frac{N_d }{2}}}W(p),
\end{equation}
where
\begin{equation}\label{w2}
 W(p)= 2^{N_d -1} ~^{N_d \otimes}\la 0| \prod_{i}((1-p)I +
p X_{i})|K\ra.
\end{equation}

We now show that the $W(p)$ can be interpreted as an important
quantity in a noisy TC state. To this end, we consider a
probabilistic bit-flip noise on the TC state where an $X$ operator
is applied on each qubit with a probability of $p$. We consider
density matrix of the model after applying a quantum channel
corresponding to the bit-flip noise.  Such a noise leads to
different patterns of errors constructed by $X$ operators on
qubits and we denote such an error by $\hat{\mathcal{E}}(X)$. The
probability of such an error is equal to $W_{\mathcal{E}}(p)
=p^{|\mathcal{E}|} (1-p)^{N_d -|\mathcal{E}|}$ where
$|\mathcal{E}|$ refers to the number of qubits which have been
affected by the noise and $N_d $ is the total number of qubits.
The effect of bit-flip noise on an arbitrary N-qubit quantum
state, denoted by density matrix $\rho$, can be presented by a
quantum channel in the following form:
\begin{equation}\label{p1}
\Phi(\rho)=\sum_{\mathcal{E}}W_{\mathcal{E}}(p) \hat{\mathcal{E}}(X) \rho \hat{\mathcal{E}}(X)
\end{equation}

Now, we come back to the relation of $W(p)$ in Eq.(\ref{w2}) and
suppose $p$ as the probability of bit-flip noise. If we expand $
\prod_{i}((1-p)I +p X_{i})$ in this equation, we will have a
superposition of all possible errors with the corresponding
probability in the following form:

\begin{equation}
\prod_{i}((1-p)I +p X_{i})=\sum_{\mathcal{E}}W_{\mathcal{E}}(p) \hat{\mathcal{E}}(X).
\end{equation}

On the other hand, the toric code state on the dual lattice is in
the form of $ |K\ra
=\frac{1}{\sqrt{2^{|f_d|}}}\prod_{f}(I+B_f)|0\ra^{\otimes N_d}$
where the operator $\prod_{f}(I+B_f)$ is also a superposition of all possible $X$-type loop operators if we interpret the Identity operator as a loop operator with a zero length (an interpretation that will be supposed in the following of the paper). Therefore, it will be easy
to see that $W(p)$ is equal to the probability of generating loops
in the noisy TC state. By this fact, the Eq.(\ref{ds}) is a
relation between partition function of Ising model on a 2D square
lattice and the probability of generating loops in the noisy TC
state.

\section{phase transition in a noisy TC model}\label{s3}

Eq.(\ref{ds}) shows that there is a relation between 2D Ising
model and a noisy TC state. Since in the 2D Ising model it is
well-known that there is phase transition at a critical
temperature $T_{cr}$, one can ask if there is a phase transition
in the noisy TC model at a corresponding probability of $p_{cr}$.
In this section, we consider this problem where we start with
magnetization of 2D Ising model and find its quantum analogue in
the noisy TC model. Then we show that such a quantum analogue is
in fact an order parameter in the noisy TC model which
characterizes a transition from a coherent to a non-coherent
phase.

We start by considering the mean value of the product of two
arbitrary spins in 2D Ising model, i.e. the correlation function.
The correlation function is formally given by:
\begin{equation}\label{wp5}
\la \sigma_m \sigma_n \ra =\frac{1}{\mathcal{Z}}\sum_{\{\sigma_i \}}\sigma_m \sigma_n e^{\beta J\sum_{\la i ,j \ra}\sigma_i \sigma_j}
\end{equation}

\begin{figure}[t]
\centering
\includegraphics[width=8cm,height=4.5cm,angle=0]{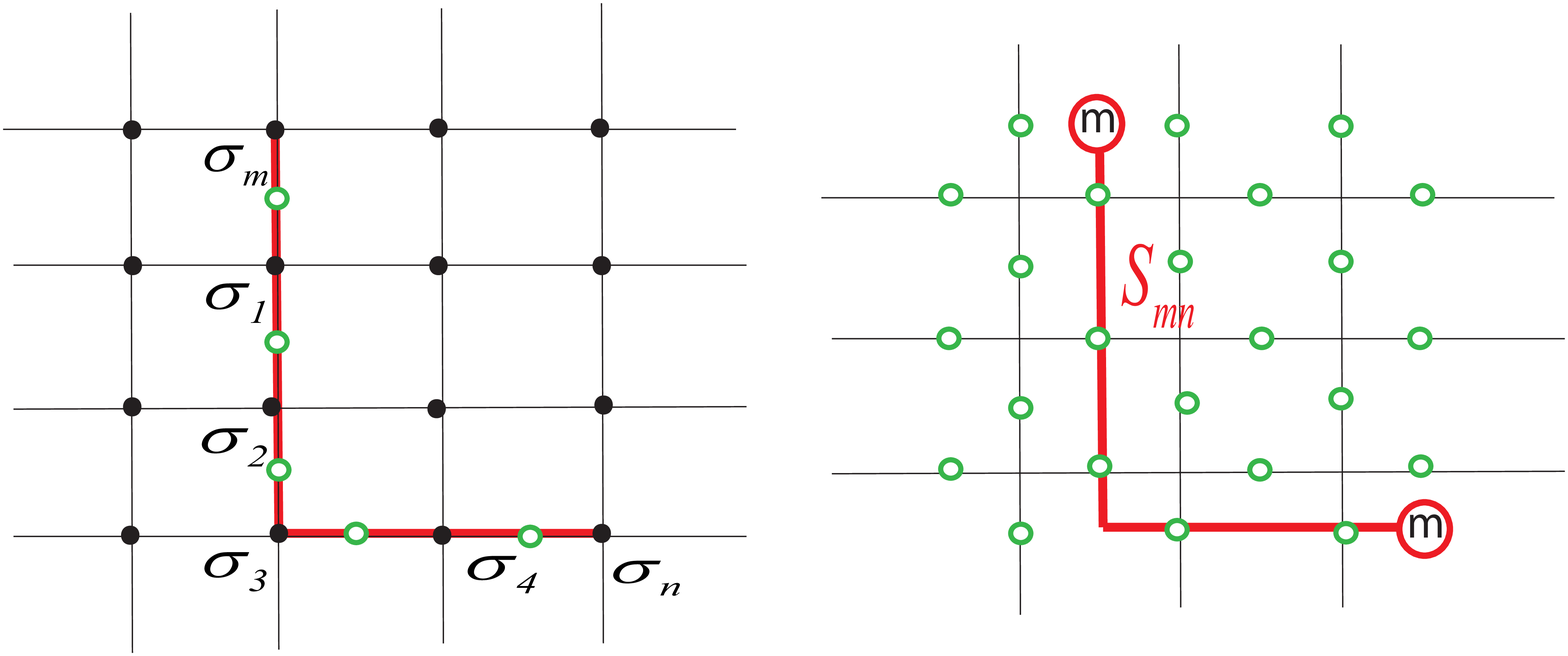}
\caption{(Color online) Left: There is a chain of spins
$\sigma_1$, $\sigma_2$, $\sigma_3$ and $\sigma_4$ between two
spins $\sigma_m$ and $\sigma_n$ . Right: after change of variables
and converting to quantum language, there will be a string
operator corresponding to the initial string $S_{mn}$.} \label{k5}
\end{figure}
Now, let us consider two particular spins as is shown in
Fig.(\ref{k5}). Then we consider an arbitrary string of spins
denoted by $S_{mn}$ between two spins $\sigma_m$ and $\sigma_n$
which we denote by $\sigma_1 ,..., \sigma_4$. Since $\sigma_i ^2
=1$, we can write $\sigma_m \sigma_n = (\sigma_m \sigma_1)(
\sigma_1 \sigma_2) (\sigma_2 \sigma_3) (\sigma_3 \sigma_4)(
\sigma_4 \sigma_n)$. Next, we write Eq.(\ref{wp5}) in the
following form:
$$
\la \sigma_m \sigma_n \ra =
$$
\begin{equation}\label{mp}
 \frac{\sum_{\{\sigma_i
\}} (\sigma_m \sigma_1)( \sigma_1 \sigma_2) (\sigma_2 \sigma_3)
(\sigma_3 \sigma_4)( \sigma_4 \sigma_n) e^{\beta J\sum_{\la i ,j
\ra}\sigma_i \sigma_j}}{\mathcal{Z}}
\end{equation}
By this new form, we can use edge spins that we defined in the
previous section to rewrite the above relation in terms of edge
spins $\xi_l$. We will have:
\begin{equation}\label{bp}
\la \sigma_m \sigma_n \ra =\frac{1}{\mathcal{Z}}\sum_{\{\xi_l \}} [\prod_{l\in S_{mn}}\xi_l ] ~~e^{\beta J\sum_{l}\xi_l} \prod_{f}\delta(\prod_{l\in \partial f}\xi_l)
\end{equation}
where $l\in S_{mn}$ refers to all edges belonging to the string of
$S_{mn}$. Similar to procedure that we performed for partition
function of 2D Ising model, we can re-write the above relation in
a quantum language where we replace edge spins with the Pauli
matrices $Z$ and by using definition of TC states, we will have:
\begin{equation}\label{vp}
\la \sigma_m \sigma_n \ra =\frac{\la \alpha | \prod_{l\in S_{mn}}Z_l |K\ra}{\la \alpha |K\ra}=\frac{\la \alpha | S_{mn}^z |K\ra}{\la \alpha |K\ra}
\end{equation}
where $S_{mn}^z$ is a $Z$-type string operator between vertices of
$m$ and $n$ and $S_{mn}^z |K\ra$ is an excited state including two
flux anyons in vertices of $m$ and $n$.

The above process can also be applied for mean value of an
arbitrary spin of the 2D Ising model denoted by $\la \sigma_i \ra$
which is in fact the order parameter of this model. Since we
considered a 2D Ising model with an open boundary condition where
all spins in the boundary of the lattice are fixed to the value of
$+1$, expectation value of an arbitrary spin $\sigma_i$ is in fact
equal to correlation function between that spin and another spin
on the boundary of the lattice denoted by $\sigma_0 =+1$.
Therefore, we can use the above formalism for the correlation
function to find a quantum formalism for $\la \sigma_i\ra$.
\begin{figure}[t]
\centering
\includegraphics[width=8cm,height=4.5cm,angle=0]{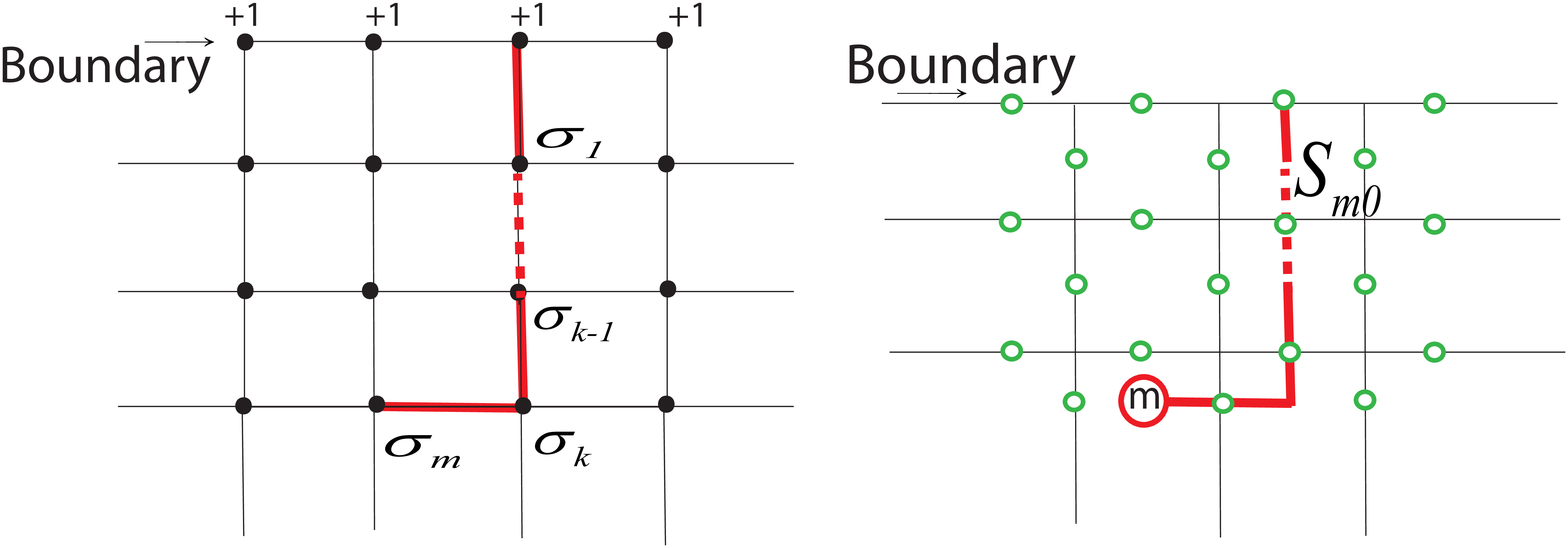}
\caption{(Color online) Left: There is a chain of spins between
boundary of the lattice and spin $\sigma_m$. Right: After change
of variables and converting to the quantum language, there will be
a string operator corresponding to the string of $S_{0m}$ which
corresponds to a flux anyon in the TC model with an open
boundary.} \label{k6}
\end{figure}
As we show in Fig.(\ref{k6}), we consider a string of spins
$\sigma_1,...,\sigma_k$ between a particular spin $\sigma_m$ and a
spin $\sigma_0 =+1$ on the boundary of the lattice. In this way,
we will have $\sigma_m=(\sigma_m \sigma_1)( \sigma_1 \sigma_2)
...(\sigma_{k-1} \sigma_{k})(\sigma_k \sigma_0) $ and the order
parameter will be in the following form:
$$\la \sigma_m  \ra =$$
\begin{equation}\label{rp}
\frac{\sum_{\{\sigma_i \}} (\sigma_m \sigma_1)( \sigma_1 \sigma_2)...( \sigma_{k-1}\sigma_k)(\sigma_k \sigma_0) e^{\beta J\sum_{\la i ,j \ra}\sigma_i \sigma_j}}{\mathcal{Z}}
\end{equation}
Then we use a change of variable to edge spins $\xi_l$ and finally
the above equation is re-written in a quantum language in the
following form:
\begin{equation}\label{pp}
\la \sigma_m  \ra =\frac{ \la \alpha | \prod_{l\in S_{m0}}Z_l |K\ra}{\la \alpha |K\ra}=\frac{\la \alpha | S_{m0}^z|K\ra}{\la \alpha |K\ra}
\end{equation}
where $S_{m0}$ refers to a string with one of its end-points on
the boundary of the lattice and the other on the face of $m$.
Furthermore, since $S_{m0}^z$ is an Z-type string operator, the
quantum state of $ S_{m0}^z |K\ra$ is an excited state of the TC
model with only one flux anyon in one of the end-points of the
string.

In order to relate the above result to a noisy TC model, we use
the transformation from Boltzmann weights in the $|\alpha \ra$ to
probability of noise $p$ according to $p=\frac{e^{-2\beta
J}}{1+e^{-2\beta J}}$. In this way, according to Eq.(\ref{ds}) in
the previous section, it is clear that the denominator in
Eq.(\ref{pp}) is the same as the partition function up to a factor
$\sqrt{2^{|f_d|}}$ and will in fact be equal to
$\frac{W(p)}{\sqrt{2^{|f_d|}}[p(1-p)]^{\frac{N_d}{2}}}$.
Furthermore, we need to find an interpretation for the numerator
of Eq.(\ref{pp}). To this end, we note that after transformation
to probability of noise $p$, the numerator is written in the
following form:
\begin{equation}
\la \alpha | S_{m0}^z |K\ra= \frac{^{ N\otimes}\la 0| (p X+(1-p)I)^{\otimes N} S_{m0}^z|K\ra}{\sqrt{2^{|f_d|}}[p(1-p)]^{\frac{N}{2}}}
\end{equation}
where we have replaced the $N_d$ by $N$ for simplifying the notation. As noted previously, $ (p X+(1-p)I)^{\otimes N}$ is equal to
superposition of all possible bit-flip errors. On the other hand,
in the $S_{m0}^z|K\ra=S_{m0}^z \prod_{f}(I+B_f)|0\ra ^{\otimes
N}$, $\prod_{f}(I+B_f)$ is equal to superposition of all X-type
loop operators with the same weight $+1$. However, the string
operator $S_{m0}^z$ does not commute with $X$-type loop operators
that cross the string of $S_{m0}$  odd number of times. Therefore,
one can conclude that in the state of $S_{m0}^z|K\ra$ we will have
a superposition of all $X$-type loop operators with two different
weights $+1$ for loops that cross the string $S_{m0}$ even number
of times and $-1$ for loops that cross the $S_{m0}$ odd number of
times. By this fact, the numerator will be related to the total
probability, denoted by $W_+ (p)$, that noise leads to loops with
even crossings with the $S_{m0}$  minus total probability, denoted
by $W_- (p)$, that noise leads to loops with odd crossings with
$S_{m0}$. Finally, Eq. (\ref{pp}) will be in the following form:
\begin{equation}\label{eqs}
\la \sigma_m \ra =\frac{W_+ (p)-W_- (p)}{W_+ (p) + W_- (p)}
\end{equation}
where, we have replaced $W(p)$ in the denominator by $W_+ (p) +
W_- (p)$ and we have removed a factor of $
\frac{1}{\sqrt{2^{|f_d|}}[p(1-p)]^{\frac{N}{2}}}$ from
denominator and nominator.

In the above relation on the left hand side, we have the order
parameter of the 2D Ising model and the right hand is a quantity
related to the noisy toric code state. Now we need to find a
physical interpretation for this quantity. Specifically, it will
be interesting to show that the right hand side is an order
parameter in the noisy TC which characterizes a type of phase
transition. To this end, we come back to the noisy model and we
consider the effect of the bit-flip noise on a particular initial
state. Suppose that the initial state is an eigenstate of string
operator $S_{m0}^z$. It is simple matter to check that such a
state will be in the following form:
\begin{equation}
|\psi_{+} \ra =\frac{1}{\sqrt{2}}(|K\ra + S_{mo}^z |K\ra)
\end{equation}
The above state is in fact a coherent superposition of a vacuum
state $|K\ra$, where there is no anyon, and a two anyon state
$S_{m0}^z |K\ra$ where one lives on the boundary $0$ and the other
on the face of $m$. Our main purpose is to
investigate the decoherence process of this coherent superposition
as a result of noise. We expect that by increasing the probability
of the bit flip noise, the initial state converts to a complete
mixture of $|K\ra \la K|$ and $S_{m0}^z |K\ra \la K| S_{m0}^z$.
However, the actual trend, as a function of noise probability,
that such a transition to decoherence occurs is an important
consideration.  For example, is the transition a gradual one or is
there a second-order transition? How can one characterize such a
transition.  Next, we set out to show that such a transition is in
fact sharp and can be characterized by an order parameter which
measures the amount of coherence in the system.

\begin{figure}[t]
\centering
\includegraphics[width=8cm,height=3.5cm,angle=0]{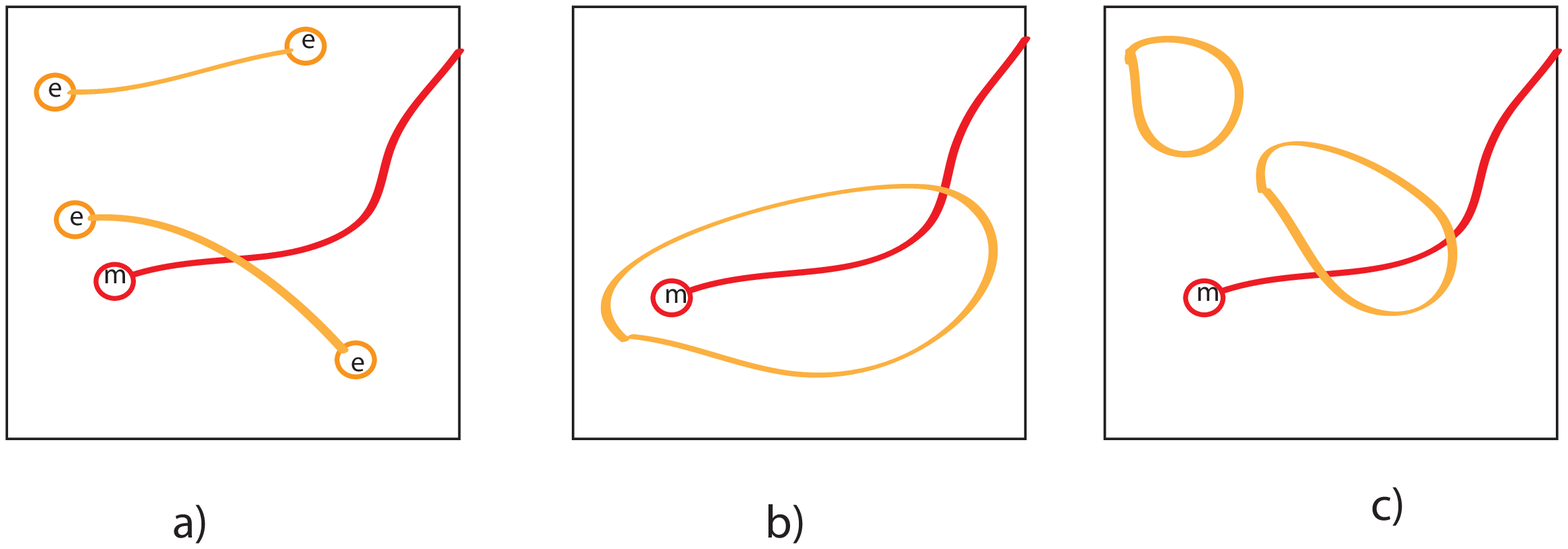}
\caption{(Color online) Red string refers to the string operator
$S_{m0}^z$ with a flux anyon in the end-point. a) The first set of
errors $E_1$ which correspond to open strings of $X$ operators
with two charge anyons in the end-points. b) The second set of
errors $E_2$ which correspond to loops of $X$ operators which
cross the $S_{m0}$ odd number of times. C) The third set of errors
$E_3$ which correspond to loops of $X$ operators which cross the
$S_{m0}$ even number of times. } \label{k7}
\end{figure}
In order to consider the above decoherence process, we divide all
errors $\mathcal{E}(X)$ in Eq.(\ref{p1}) to three parts. By the
fact that each error of $\mathcal{E}(X)$ can be represented as a
pattern of string operators $X$ on the lattice, we consider three
kinds of strings which are schematically shown in Fig.(\ref{k7}).
The first are open strings where there are charge anyons in the
endpoints of those strings. The second are closed strings (loops)
that cross the string of $S_{m0}$ odd number of times and the
third are closed strings that cross the string of $S_{m0}$ even
number of times. It is simple to check that the effect of errors
corresponding to the open strings, denoted by $E_1$, on the state
of $|\psi_+ \ra$ leads to generation of charge anyons on endpoints
of open strings and takes the initial state to other excited
states. The effect of the second kind of errors, denoted by $E_2$,
is interesting where it takes $|\psi_+\ra$ to
$\frac{1}{\sqrt{2}}(|K\ra - S_{mo}^z |K\ra)$, denoted by
$|\psi_-\ra$ which is orthogonal to $|\psi_+\ra$, because loop
operators of the second type anti-commute with $S_{m0}^z$. Finally
the effect of the third kind of errors, denoted by $E_3$, is
trivial as it takes $|\psi_+\ra$ to $|\psi_+\ra$. In this way,
Eq.(\ref{p1}), when we insert $\rho =|\psi_+ \ra \la \psi_+ |$,
can be written in the following form:

$$ \Phi(|\psi_+  \ra \la \psi_+ |)=\sum_{\mathcal{E}\in E_1 }W_{\mathcal{E}}(p) |\psi_{\mathcal{E}} \ra \la \psi_{\mathcal{E}}| $$
\begin{equation}\label{p10}
+ W_{-}(p) |\psi_{-}\ra \la \psi_-| + W_{+}(p) |\psi_{+}\ra \la \psi_+ |
\end{equation}
where
$$ W_{-}(p)= \sum_{\mathcal{E}\in E_2 }W_{\mathcal{E}}(p),$$
$$ W_{+}(p)=\sum_{\mathcal{E}\in E_3 } W_{\mathcal{E}}(p)$$
and
$$  |\psi_{\mathcal{E}} \ra = \mathcal{E}(X) |\psi \ra , $$
and $E_1 $, $E_2$ and $E_3$ refer to the above three kinds of
errors, respectively. $W_+$ and $W_-$ are total probability of
generating loops which cross the $S_{m0}$ even and odd number of
times, respectively. In order to find a better interpretation for
the above state, note that in the final quantum state there is a
mixture of $|\psi_+\ra$ with probability of $W_+ (p)$ and $|\psi_-
\ra$ with probability of $W_- (p)$. In other words, while the
state of $|\psi_+\ra = \frac{1}{\sqrt{2}}(|K\ra + S_{mo}^z |K\ra)$
is a coherent superposition of states $|K\ra$ and $S_{m0}^z
|K\ra$, the effect of bit-flip noise has led to generating a
non-coherent mixture of them which can be represented in the
corresponding subspace in the following form:
\begin{equation}\label{lk}
 \frac{1}{2}\left(
  \begin{array}{cc}
    (W_+ +W_- ) & (W_+ -W_- ) \\
    (W_+ -W_- ) & (W_+ +W_-) \\
  \end{array}
\right)
\end{equation}
Now, we need to define a parameter to characterize the amount of
decoherency in the above mixture in terms of $p$. According to the
Eq.(\ref{lk}), the following parameter is a well-defined measure
for the above decoherency:
\begin{equation}\label{oit}
O (p)=\frac{W_+ (p) -W_- (p)}{W_+ (p) +W_- (p)}
\end{equation}
\begin{figure}[t]
\centering
\includegraphics[width=9cm,height=6cm,angle=0]{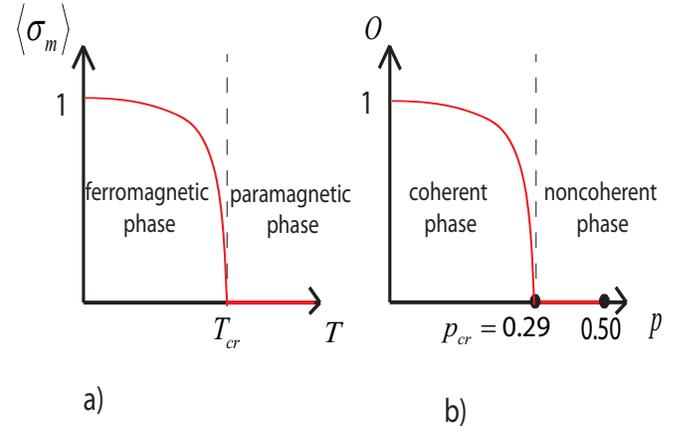}
\caption{(Color online) a) A schematic representation of the
well-known diagram of the order parameter in 2D Ising model where
the magnetization goes to zero at $T_{cr}$. b) By comparing
Eq.(\ref{eqs}) and Eq.(\ref{oit}), the expectation value of the
string operator $S_{m0}^z$ shows the same behavior with a phase
transition at $p_{cr}=0.29$. } \label{k8}
\end{figure}
This quantity can be also interpreted as the expectation value of
$S_{m0}^z$ in a subspace which is spanned by $|\psi_+\ra$ and
$|\psi_-\ra$. For example, when $W_+ =W$ and $W_- =0$ the order
parameter $O$ is equal to $1$  which indicates a coherent
superposition and when $W_+ =W_- $, it is equal to $0$
indicating a complete mixture where Eq.(\ref{lk}) becomes proportional to the
Identity operator. Now, referring to Eq.(\ref{eqs}), we see that
$O$ is identical to the order parameter of the 2D Ising model. On
the other hand, it is well-known that the order parameter in 2D
Ising model characterizes the nature of ferromagnetic phase
transition where at a critical temperature $T_{cr}$ system shows a
second-order phase transition from ordered phase $\la \sigma_m \ra \neq 0$ to
an disordered phase $\la \sigma_m \ra =0$. Therefore, by using
Eq.(\ref{oit}), it is concluded that there is a phase transition
in the noisy TC state where a relatively coherent state, $O\neq 0
$ goes to a complete mixture, $O=0$, at a well defined (and
relatively large) noise value of $p_{cr}$ which is easily
calculated from the critical temperature of the 2D Ising model to
be $p_{cr}=0.29$.

The picture that emerges is very interesting.  One would expect
that increasing bit-flip noise on a coherent superposition would
lead to decoherency.  However, we have shown that the system
remains relatively robust to such an effect for small values of
noise probability $p \ll p_{cr}$ and that the transition can be
characterized by an order parameter which shows a second-order phase transition
to decoherency at a relatively large value of noise value.
Fig.(\ref{k8}) schematically shows such a behavior as increasing $p$ from zero
to half leads to decoherency at the value of $p_{cr}=0.29$. This
interesting and unexpected property indicates a \emph{robust coherency} which might be related to topological order of the TC
state, a point that we will come back to in the Sec.(\ref{s5}).

\section{Susceptibility and critical stability at phase transition}\label{s4}

Although the existence of a phase transition in a physical system
is very important by itself, the critical point which separates
two different phases of the system is also a key matter which
should specifically be considered. Therefore, we intend to look
for other consequences of criticality on the classical side for
the quantum model. In this section we consider this problem and
specifically show that the ground state of the TC model under the
noise displays an interesting behavior precisely at the critical
point $p_{cr}$.

We consider an important issue that has been previously emphasized
in the general context of CSS states, i.e. relative \emph{critical
stability}, which occurs at $p_{cr}$ \cite{zare18}. Here, we
investigate such a concept in terms of a susceptibility to noise
which is defined for an initial state affected by a bit-flip
noise. To this end, we use a familiar quantity in the quantum
information theory called fidelity. In other word, if we consider
the state of $|K\ra$ as an initial state, the fidelity of this
state and the final state after applying noise will be in the
following form:
\begin{equation}\label{q2}
F(p)=\la K |\Phi(\rho) |K\rangle
\end{equation}
where $\Phi(\rho)$ is the final state after applying the bit-flip
channel to the initial state $\rho =|K\ra \la K|$. Now it is clear
that if $F(p)$ is small (large) it means that susceptibility to
noise is high (low) because the final state is very different from
(similar to) the initial state. In other words, there is an
inverse relation between susceptibility to noise and fidelity.
Therefore, we define the following quantity to measure the
susceptibility:
\begin{equation}\label{sfg}
\chi (p)=-Log(F(p)).
\end{equation}
It will be interesting to give a physical picture to this
quantity. To this end, we interpret $F(p)$ in an anyonic picture
for the TC state. In fact, when a bit-flip noise is applied to a
qubit of the lattice, it generates two charged anyons. Therefore, the effect
of probabilistic bit-flip noise on all qubits can lead to two events,
generation of pair anyons and walking anyons in the lattice. It is
clear that as long as there is an anyon in the lattice, the system
is in an excited mode and fidelity is zero, i.e. susceptibility is
infinite. The only possible way that the system comes back to the
ground state is that anyons fuse to each other and annihilate. In
other words, anyons must generate complete loops in the lattice.
This consideration seems to indicate that $F(p)$ should be equal
to probability of generating loops which is the same as $W(p)$.

In order to explicitly prove that $F(p)=W(p)$, we use definition
of Eq.(\ref{q2}) for $F(p)$. On one hand, we know that
$|K\ra=\frac{1}{\sqrt{{2}^{|f|}}}\prod_{f}(I+B_f )|0\rangle
^{\otimes N}$ and $\prod_{f}(I+B_f )$ is equal to summation of all
possible $X$-type loop operators in the lattice. On the other
hand, $\Phi (\rho)$ in Eq.(\ref{q2}) is equal to $\sum_{\mathcal{E}}W_{\mathcal{E}}(p) \hat{\mathcal{E}}(X) \rho \hat{\mathcal{E}}(X)$ which is equal to a
summation of all possible $X$-chains with the corresponding
probability. In this way, when such a summation inserts between
two states $|K\ra$ in the Eq.(\ref{q2}), all $X$-chains in the
summation lead to error and convert the inner product to zero,
except for the $X$-chains corresponding to loops which convert the
inner product to $1$. Therefore, there will be a summation of
probability of generating loops and it is equal to $W(p)$.

Now, we note that $W(p)$ was related to the partition function of
2D Ising model according to Eq.(\ref{ds}) and since $F(p)=W(p)$,
we have:
\begin{equation}\label{ds2}
\mathcal{Z}=\frac{1}{[p(1-p)]^{\frac{N}{2}}}F(p),
\end{equation}
Therefore, one can find the fidelity in the form of
$[p(1-p)]^{\frac{N}{2}} \mathcal{Z}$. Since $p\leq \frac{1}{2}$,
it is concluded that $[p(1-p)]^{\frac{N}{2}}\rightarrow 0$ for
large $N$. In this way, it seems that $F(p)$ should be zero for
any non-zero value of $p$ and therefore susceptibility is infinite for any
generic noise probability.  However, at the critical point, the partition
function displays a non-analytic behavior, where fluctuations
become relevant. A fluctuation corrections to partition function
can be written as \cite{pathr}:
\begin{equation}
\mathcal{Z}\approx\exp(-\beta A)\sqrt{2\pi kT^2C_v},
\end{equation}
where $A$ is the Helmholtz free energy and $C_v$ is the heat
capacity which behaves as $C_v \sim \mid T-T_{cr}\mid^{-\alpha}$,
near the critical point. Fidelity is clearly equal to unity for
zero noise (or temperature), but it is also a strongly decreasing
function of $p$ as can be seen from Eq.(\ref{ds2}).  However, the
divergence of heat capacity at the critical point will cause a
relative increase of the value of fidelity (or relative decrease of susceptibuility) as the critical point
is approached, thus leading to a \emph{relative stability}
\cite{zare18}.

There are two points about critical stability that should be
emphasized. The first is that, the concept of susceptibility that
we defined here is not the usual susceptibility in classical phase
transitions. For example in the Ising model, heat capacity can be
interpreted as susceptibility of the system to an infinitesimal
change of temperature. However, in our case, the susceptibility
measures stability of the initial state to the whole of the noise
not an infinitesimal change of the probability of the noise.  In
other words, our susceptibility is not defined as a derivative,
but as the response of the ground state to a noise probability of
finite value $p$.

As the second point, we emphasize that the critical probability of
$p_{cr}$ should not be regarded as a threshold for stability of
the toric code state. Contrary, it is a relative stability that
occurs only at a particular noise probability, $p_{cr}$. It is
different from the role of $p_{cr}$ in the previous section where
it was a threshold for maintaining coherency of a particular
initial state.

A physical picture may help to clarify what is happening in both
situations. As $p$ increases the possibility of forming larger
loops also increases. It is at $p_{cr}$ where the possibility of
having loops of the order of system size, $N$, appear.  This is
related to the fact that correlation length diverges at the
critical point. This system size loops are prime candidates for
increasing the value of $W_{-}$, as they are prime candidates for
crossing the string operators once, when compared to smaller
strings which typically don't cross or cross twice, see Fig.(\ref{k7}).
This explains how the order parameter suddenly drops to zero at
the critical point as a significant $W_{-}$ cancels out the
already reduced $W_{+}$.  Also note that for $p>p_{cr}$, both
values of $W_+$ and $W_-$ are significantly small and equal
leading to zero order parameter. On the other hand, the emergence
of such system size loops which can occur only at $p_{cr}$, give a
relative increase to the value of $W(p)= W_+ +W_-$, thus
explaining the relative stability near (at) the critical point.

\section{discussion}\label{s5}
Although some time has passed since the  introduction of quantum
formalism for the partition function of classical spin models, it
seems that such mappings are richer than what had been previously
supposed. In particular, here we studied a neglected aspect of
such mappings related to the phase transition on the classical
side. In particular, we observed the existence of a sharp
transition from a coherent phase to a mixed phase in the noisy TC
model as well as a relative critical stability at the transition
point. These seem like important physical properties which might
find relevance in the applications of quantum states in general.
As a closing remark, we would like to emphasize that the large
value of the critical noise $P_{cr}=0.29$ can be interpreted as a
robust coherency of the initial state against bit-flip noise. On
the other hand, since string operators and loop operators in the
TC model correspond to processes of generating and fusing anyons,
it seems that the robust coherency is in fact related to anyonic
properties. This point beside a relative stability at the critical
noise support a conjecture that such behaviors might be related to
topological order of the TC state. Since, the critical stability
has also been observed in other topological CSS states
\cite{zare18}, it will be interesting to consider the existence of
the robust coherency in such models, a problem we intend to
address in the future studies.

\section*{Acknowledgement}
We whould like to thank A. Ramezanpour for their valuable comments on this paper.


\begin{thebibliography}{99}
\bibitem{Geraci2008}
J. Geraci, D. A. Lidar, ``On the exact evaluation of certain instances of the Potts partition function by quantum computers," Commun. Math. Phys. 279, 735 (2008).
\bibitem{Lidar1997}
D. A. Lidar, O. Biham, ``Simulating Ising spin glasses on a quantum computer," Phys. Rev. E 56, 3661 (1997).
\bibitem{Geraci2010}
J. Geraci, D. A. Lidar, ``Classical Ising model test for quantum circuits," New J. Phys. 12, 75026 (2010).
\bibitem{Somma2007}
R. D. Somma, C. D. Batista, G. Ortiz, ``Quantum approach to classical statistical mechanics," Phys. Rev. Lett. 99, 030603 (2007).
\bibitem{Verstraete2006}
F. Verstraete, M. Wolf, D. Perez-Garcia, and J. Cirac, ``Criticality, the area law, and the computational power of projected entangled pair states," Phys. Rev. Lett. 96 (2006).
\bibitem{Dennis2002}
E. Dennis, A. Kitaev, A. Landahl, and J. Preskill, ``Topological quantum memory," J. Math. Phys. 43, 4452 (2002).
\bibitem{Katzgraber2009}
H. G. Katzgraber, H. Bombin, M. A. Martin-Delgado, ``Error threshold for color codes and random three-body Ising models," Phys. Rev. Lett. 103, 090501 (2009).
\bibitem{ent2006}
S. Popescu, A. J. Short, and A. Winter, ``Entanglement and the
foundations of statistical mechanics." Nature Physics 2, no. 11
(2006): 754.
\bibitem{mont2010}
 A. Montakhab, A. Asadian, ``Multipartite entanglement and quantum phase transitions in the one-, two-, and three-dimensional transverse-field Ising
model", Phys. Rev. A 82, 062313 (2010).
\bibitem{eis17}
C. Gogolin, and J. Eisert, ``Equilibration, thermalisation, and the emergence of statistical mechanics in closed quantum systems." Reports on Progress in Physics 79, no. 5 (2016): 056001.
\bibitem{termo}
J. Goold, M. Huber, A. Riera, L. del Rio, and P. Skrzypczyk, ``The role of quantum information in thermodynamics—a topical review." Journal of Physics A: Mathematical and Theoretical 49, no. 14 (2016): 143001.
\bibitem{Nest2007}
M. Van den Nest, W. D\"{u}r, H. J. Briegel, ``Classical spin models and the quantum-stabilizer formalism," Phys. Rev. Lett. 98, 117207 (2007).

\bibitem{algor}
M. Van den Nest, W. D\"{u}r, R. Raussendorf, and H. J. Briegel, "Quantum algorithms for spin models and simulable gate sets for quantum computation." Physical Review A 80, no. 5 (2009): 052334.
\bibitem{durmar}
W. D\"{u}r, M. Van den Nest, "Quantum simulation of classical thermal states." Physical review letters 107, no. 17 (2011): 170402.
\bibitem{gemma}
G. De las Cuevas, ``A quantum information approach to statistical mechanics." Journal of Physics B: Atomic, Molecular and Optical Physics 46.24 (2013): 243001.
\bibitem{Cuevas2011}
G. De las Cuevas, W. D\"{u}r, M. Van den Nest and M. A. Martin-Delgado, ``Quantum algorithms for classical lattice models," New J. Phys. 13:093021 (2011).
\bibitem{mqc}
R. Raussendorf, H. J. Briegel, ``A one-way quantum computer." Physical Review Letters 86, no. 22 (2001): 5188.
\bibitem{mbqc}
H. J. Briegel, D. E. Browne, W. D\"{u}r, R. Raussendorf, M. Van den Nest, ``Measurement-based quantum computation." Nature Physics, 5(1), 19 (2009).
\bibitem{Bravyi2007}
S. Bravyi, R. Raussendorf, ``Measurement-based quantum computation with the toric code states," Phys. Rev. A 76, 022304 (2007).
\bibitem{Bombin2008}
H. Bombin, M. A. Martin-Delgado, ``Statistical mechanical models and topological color codes," Phys. Rev. A 77, 042322 (2008).
\bibitem{Nest2008}
M. Van den Nest, W. D\"{u}r, H. J. Briegel, ``Completeness of the classical 2D Ising model and universal quantum computation," Phys. Rev. Lett. 100, 110501 (2008).
\bibitem{Vahid2012b}
V. Karimipour. M. H. Zarei, ``Algorithmic proof for the completeness of the two-dimensional Ising model," Phys. Rev. A 86, 052303 (2012).
\bibitem{Cuevas2009}
G. De las Cuevas, W. D\"{u}r, H. J. Briegel, and M. A. Martin-Delgado, ``Unifying all classical spin models in a Lattice Gauge Theory," Phys. Rev. Lett. 102, 230502 (2009).
\bibitem{xu}
Y. Xu, G. De las Cuevas, W. D\"{u}r, H. J. Briegel, M. A. Martin-Delgado, ``The U (1) lattice gauge theory universally connects all classical models with continuous variables, including background gravity," J. Stat. Mech. 1102:P02013 (2011).
\bibitem{Vahid2012}
V. Karimipour, M. H. Zarei, ``Completeness of classical $\phi^4$ theory on two-dimensional lattices," Phys. Rev. A 85, 032316 (2012).
\bibitem{yahya}
Mohammad Hossein Zarei, Yahya Khalili, ``Systematic study of the completeness of two-dimensional classical φ4 theory," Int. J. Quantum Inform. 15, 1750051 (2017).
\bibitem{science}
G. De las Cuevas, T. S. Cubitt, ``Simple universal models capture all classical spin physics," Science 351.6278 : 1180-1183 (2016).
\bibitem{cub}
T. S. Cubitt, A. Montanaro, and S. Piddock, ``Universal quantum hamiltonians." Proceedings of the National Academy of Sciences 115, no. 38 (2018): 9497-9502.
\bibitem{zare18}
M. H. Zarei, A. Montakhab, ``Dual correspondence between classical spin models and quantum CSS states," Phys. Rev. A 98, 012337 (2018).
\bibitem{stan}
H. E. Stanley, \emph{Phase transitions and critical phenomena}, Clarendon Press, Oxford, 1971.
\bibitem{pathr}
Pathria, R. K. \emph{Statistical Mechanics}, International Series in Natural Philosophy Volume 45, Pergamon Press, Oxford, UK, 1986.
\bibitem{Kitaev2003}
A. Y. Kitaev, ``Fault-tolerant quantum computation by anyons," Ann. Phys. (N.Y.) 303, 2 (2003).
\bibitem{wen}
X.-G. Wen, and Q. Niu, ``Ground-state degeneracy of the fractional quantum Hall states in the presence of a random potential and on high-genus Riemann surfaces." Physical Review B 41, no. 13 (1990): 9377.
\bibitem{wen2}
X.-G. Wen, ``Topological orders and edge excitations in fractional quantum Hall states." Adv. Phys. 44, 405 (1995).
\bibitem{rob1}
S. Trebst, P. Werner, M. Troyer, K. Shtengel, and C. Nayak, ``Breakdown of a topological phase: Quantum phase transition in a loop gas model with tension." Phys. Rev. Lett. 98, 070602 (2007)
\bibitem{rob2}
S. Dusuel, M. Kamfor, R. Orus, K. P. Schmidt, and J. Vidal, "Robustness of a perturbed topological phase." Physical Review Letters, 106, 107203, (2011).

\bibitem{zare16}
M. H. Zarei, "Robustness of topological quantum codes: Ising perturbation." Physical Review A 91, no. 2 (2015): 022319.
\bibitem{zareiprb}
Mohammad Hossein Zarei, ``strong-weak coupling duality between two perturbed quantum many-body systems: CSS codes and Ising-like systems ,"  Phys. Rev. B 96, 165146 (2017).
\bibitem{errork}
E. Dennis, A. Kitaev, A. Landahl, and J. Preskill, ``Topological quantum memory." Journal of Mathematical Physics 43, no. 9 (2002): 4452-4505.
\bibitem{errorw}
J. R. Wootton, and J. K. Pachos, ``Bringing order through disorder: Localization of errors in topological quantum memories." Physical review letters 107, no. 3 (2011): 030503.
\bibitem{errorb}
B. J. Brown, D. Loss, J. K. Pachos, C. N. Self, and J. R. Wootton. "Quantum memories at finite temperature." Reviews of Modern Physics 88, no. 4 (2016): 045005.
\end{thebibliography}
\end{document}